# Formation of complex Langmuir and Langmuir-Blodgett films of water soluble rosebengal

S. Biswas [a, b], D. Bhattacharjee [a], R. K. Nath[b] S. A. Hussain [a]*

Department of Physics [a] and Department of Chemistry [b],

Tripura University, Suryamaninagar-799130, Tripura, INDIA

**Abstract:** This communication reports the formation of complex Langmuir monolayer at the air-water interface by charge transfer types of interaction with the water soluble N- cetyl N, N, N trimethyl ammonium bromide (CTAB) molecules doped with rosebengal (RB), with the stearic acid (SA) molecules of a preformed SA Langmuir monolayer. The reaction kinetics of the formation of RB-CTAB-SA complex monolayer was monitored by observing the increase in surface pressure with time while the barrier was kept fixed. Completion of interaction kinetics was confirmed by FTIR study. This complex Langmuir films at the air-water interface was transferred onto solid substrates at a desired surface pressure to form multilayered Langmuir-Blodgett films. Spectroscopic characterizations reveal some molecular level interactions as well as formation of microcrystalline aggregates depending upon the molar ratios of CTAB and RB within the complex LB films. Presence of two types of species in the complex LB films was confirmed by fluorescence spectroscopy.

* Corresponding author
E.mail: tuphysic@sancharnet.in
         sa_h153@hotmail.com
Phone: +91-381-2375317
Fax: +91-381-2374802

## Introduction:

In recent years, the Langmuir-Blodgett technique for the preparation of ultrathin films of various organic, metallorganic and polymeric compounds plays an increasingly important role as a means of organizing molecular materials at the microscopic level [1-3]. The Langmuir Blodgett (LB) technique has many potential applications in molecular electronics, non-linear optics and conducting thin films. The most important advantage of this technique is that the characteristics of the films can be varied by changing various LB parameters, namely, surface pressure of lifting, temperature, barrier speed, dipping speed, molar composition etc. So it is important to study different molecules having various chromophores with interesting photophysical and electrical properties, confined in the restricted geometry of the LB films to fabricate various molecular electronic devices and also to realize the basic physicochemical processes involved at the mono and multilayer films.

Recently, it has been observed that some water-soluble materials can also form stable and well-organized Langmuir monolayer at the air-water interface [4-7]. It was first observed by Vollhardt et. al. in 1997 [8] and later discussed in detail in another two papers [9, 10]. However in that work it was observed that certain soluble ampliplile show first order phase transition in the Gibbs monolayer (adsorbed layer). The adsorption kinetics as well as the phase transition was largely affected by the subphase temperature as well as also by the concentration of the amphiphile.

However, in another types of works, it has also been observed that certain water soluble cationic and anionic types of materials when interact with the amphiphilic molecules of a preformed Langmuir monolayer, adsorptions of these molecules are occurred in the monolayer and subsequently a complex Langmuir monolayer is formed [11]. Menzel et. al. reported the reaction kinetics and adsorption of poly(styrene sulfonate) and carboxymethylcellulose sodium salt onto a preformed dioctadecyldimethylammonium bromide monolayer [6, 7]. Adsorption of poly (NIPAM) on the preformed pentadecanoic acid (PDA) monolayer was demonstrated by Kawaguchi et. al. [5]. Adsorption to surfaces by electrostatic interaction and consequent formation of complex monolayer is somewhat different than the Gibbs monolayer. This adsorption to surface by electrostatic interaction plays an important role in many biological and technical processes, naturally, in blood coagulation, paper production or waste water treatment [12]. Monitoring the changes on adsorption of a sample from a solution brought into the contact of a preformed amphiphile monolayer, may allow obtaining information about the dynamics of the adsorption process. Methods like Brewster angle microscopy [8-10], ellipsometry [13], surface plasmon spectroscopy [14, 15], changes in UV-Vis absorption spectra [16], changes in surface pressure (at constant area) [4], changes in the reflectivity at the Brewster angle [17] have been used for the in-situ investigation of the adsorption process to surface. Employing the stable monolayer of ionic amphiphiles at the air-water interface is a complementary method to study the adsorption of water soluble molecules in this monolayer by electrostatic interaction. The charge transfer types of interaction between the monolayer molecules and the adsorbed species forms a complex monolayer at the air-water interface. Consequently a change in the monolayer structure is occurred. Here we report that the whole process of adsorption can be monitored by changing the monolayer surface pressure with time at the air-water interface, while the barrier is kept fixed.

The advantage of this monolayer using as model surface is that the charge density and charge mobility can be adjusted by the structure of the amphiphiles and the surface pressure applied to the monolayer.

In one of our previous studies we have investigated the adsorption process of a cationic molecule namely N- cetyl N, N, N trimethyl ammonium bromide (abbreviated as CTAB) onto stearic acid (SA) monolayer [4]. The reaction kinetics of the formation of the complex monolayer of anionic SA and cationic CTAB molecules was monitored by observing the changes in surface pressure with time, while the barrier was kept fixed.

In the present investigation, we report the adsorption process of a highly fluorescent anionic dye rosebengal (RB) doped into CTAB molecules, to the preformed stearic acid (SA) monolayer at the air-water interface.

The complex monolayer was transferred onto quartz substrates at desired surface pressure to form mono and multilayered LB films. The spectroscopic characteristics of this LB film of the complex monolayer were also studied in the light of UV-Vis absorption and steady state fluorescence spectroscopy.

**Experimental:**

CTAB and rosebengal (RB) were purchased from Loba Chemie, Mumbai, India and stearic acid (purity > 99%) purchased from Sigma Chemical Company were used as received. Chloroform used was of spectroscopic grade and its purity was checked by fluorescence spectroscopy before use. Langmuir–Blodgett film deposition instrument (Apex-2000C, India) was used to study the reaction kinetics by monitoring the increase in surface pressure with time as well as also for deposition of mono and multilayer Langmuir-Blodgett (LB) films. Triple distilled deionised water was used as subphase and the temperature was maintained at $24^0$ C. First of all a stearic acid (SA) monolayer was formed by spreading 200 $\mu l$ of dilute SA solution (0.5 mg/ml) in chloroform on the air-water interface. After sufficient time was allowed to evaporate the solvent, the barrier was compressed slowly to obtain the desired initial surface pressure.

After the desired surface pressure of SA monolayer was attained, the barrier was kept fixed, and dilute aqueous solution of CTAB doped with rose Bengal (RB) in different molar ratios of CTAB:RB, were slowly injected by a microsyrringe, to the back side of the barrier, so as not to disturb the preformed SA monolayer. After that corresponding increase in surface pressure with time was recorded.

It may be mentioned in this context that in one of the similar work by Menzel et. al. [6, 7], Fromherz method was used for bringing a preformed dioctadecyldimethylammonium bromide monolayer into contact with the subphase containing polyelectrolyte to monitor the adsorption kinetics. This method was also used by several others to study the protein adsorption [18, 19]. However the main disadvantages of the process is that during the transfer process, the contact area between the monolayer and the polyelectrolyte solution increases linearly and there is a possibility of the mixing of subphase during the transfer of the monolayer. Also the transfer speed affects the stability of the monolayer as a whole.

In another work, Kawaguchi et. al. [5] demonstrated the adsorption of poly (NIPAM) on the preformed pentadecanoic acid (PDA). They prepared a monolayer of PDA inside a circular ring shape Teflon coated barrier on the LB trough. Then the dilute solution of polyion was injected into the water on the outside of the ring shaped barrier. This technique is very similar to our technique. The possibility of the mixing of subphase as well as the disturbance of monolayer stability can be avoided by using this process. As soon as the water soluble polyion comes into contact with the preformed monolayer formed within the barrier the reaction kinetics started and the surface pressure started rising. After a certain time interval, depending on the amount and concentration of the polyions, the surface pressure becomes stable and the surface pressure vs. time curves show a flat platue like region indicating the completion of the adsorption process.

Although in this process a concentration gradient of water soluble CTAB-RB molecules is produced and time dependence of reaction kinetics cannot be predicted accurately in the steep slope region. However, completion of adsorption process can be definitely confirmed from the flat platue region. Also the geometry of the experimental setup may affect the reaction kinetics at the air-water interface. The area per molecule of SA of preformed SA monolayer is fixed for a particular surface pressure. However, depending on the area of the spread preformed SA monolayer on the air-water interface of the LB trough, there may be different concentration gradient of the water soluble CTAB-RB solution. So depending on the size of the spread SA monolayer the time required to attain the flat platue region in the surface pressure vs. time graph may vary for a fixed amount of CTAB-RB solution. To minimize this, in our experimental arrangement, for all the cases, the trough area of the spread SA monolayer was kept fixed by spreading equal amount of SA solution and keeping the barrier fixed at the same position.

Once the stable complex Langmuir monolayer was formed at the air-water interface, mono and multilayered LB films of this complex monolayer were prepared by transferring it onto quartz substrate. The details of the technique were described elsewhere [4]. The FTIR spectra were taken using a FTIR spectrophotometer (Perkin Elmer, Spectrum BX). UV-Vis absorption spectrophotometer (Perkin Elmer Lambda 25) and fluorescence spectrophotometer (Perkin Elmer LS 55) were used for spectroscopic characterizations.

**Results and Discussion:**
**Formation of RB-CTAB-SA Complex monolayer at the air-water interface:**

A schematic representation of the interaction scheme of RB-CTAB molecules with the SA molecules of the preformed SA monolayer is shown in figure 1. In figure 1a, it is shown that RB-CTAB molecules are going to the close proximity of the SA molecules. In figure 1b reaction kinetics started to form complex molecules. In figure 1c all the SA molecules are converted into complex molecules and a complex monolayer is thus formed.

Initially the monolayer was occupied by only the SA molecules and finally the monolayer was occupied by the complex molecules. Since the area per molecule of this complex is greater than the area per molecule of pure SA molecules, hence the area per molecule of the complex monolayer tends to increase. But as the barrier was kept fixed at a particular position, hence the tendency to increase the area per molecule was consequently manifested to the increase in surface pressure.

Thus the rise in surface pressure with the passage of time is an indication of reaction kinetics and consequent formation of complex monolayer of RB-CTAB-SA complex molecules at the air-water interface. The reaction scheme of RB-CTAB and CTAB-SA are shown below:

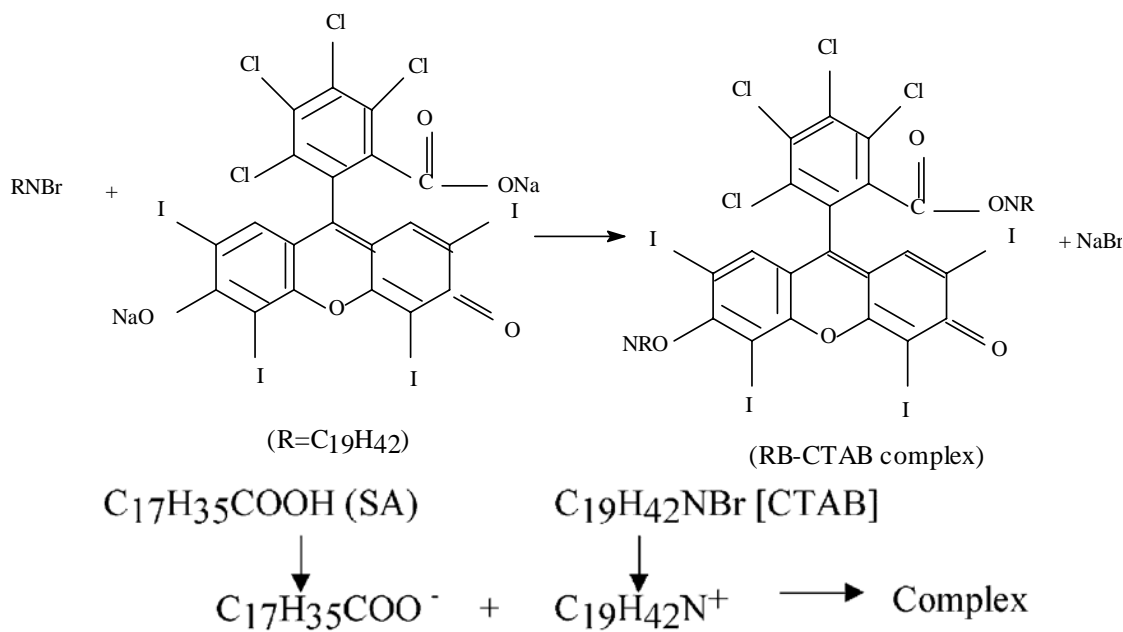

Figures 2a and 2b show the surface pressure versus time graph indicating the reaction kinetics of the RB doped CTAB molecules when interact with SA molecules of the preformed SA monolayer. The initial starting pressures were 0.5 mN/m and 15 mN/m in figure 2a and 2b respectively. RB and CTAB were mixed in different molar ratio to observe the changes in reaction kinetics. Here in all the cases, the amount of the stearic acid solution is kept fixed at 200 microlitres. The water soluble CTAB is doped with rosebengal in different molar ratios namely 90:10, 80: 20, 70:30, 60: 40, and 50: 50. Total mixed amount is 1000 $\mu l$.

From figures (2a and 2b) it is observed that the surface pressure of pure SA monolayer is parallel to time axis indicating the formation of stable Langmuir monolayer at the air-water interface. However when RB-CTAB mixed solution is injected, it is observed that the surface pressure is gradually raised with the passage of time, which clearly indicates that some interactions are taking place between the constituent molecules of CTAB and SA to form the complex monolayer. It is also observed that in the RB doped CTAB mixture, when the molar ratio of CTAB is increased, the final surface pressure is also increased. In the extreme case, when the ratio of CTAB and RB is 90:10 (900µl:100 µl), the surface pressure rises to almost 45 mN/m within a span of 3 hours from the starting surface pressure of only 0.5 mN/m. The surface pressure versus time curve shows steep rising initially but becomes flate at the end, indicating a lowering in the rate of interaction. Also the rate of increase in surface pressure becomes faster with the increase in CTAB concentration in the added solution.

It is interesting to mention in this context that there are several other molecules viz. poly (NIPAM) [5], carboxymethyl cellulose sodium salt [6], lysozyme [20] etc. show such initial steep rising and flat plateau like region at the end of the reaction kinetics. In one of our previous works [4] the reaction kinetics of pure CTAB with SA monolayer was reported. In the present observation, it is interesting to note that the initial speed of surface pressure rise is greater when RB is doped along with CTAB. This may be due to the fact that the doping element rosebengal (RB) is just getting squeezed in between the CTAB-SA complex at the air-water interface. This has been explained by fluorescence spectroscopic studies discussed later.

In order to check whether all SA molecules in the preformed SA monolayer interact with CTAB, we have taken the FTIR spectra of pure SA LB film as well as also CTAB-SA mixed LB films on silicon substrate. CTAB-SA mixed LB films were deposited at the surface pressure when flat platue like region is achieved. The FTIR spectra are shown in figure 3. The FTIR spectrum of SA LB film exhibits the characteristics strong bands at about 1710 cm$^{-1}$ which is due to the stretching vibration of carbonyl group [21]. This band is the indicative of acid form of SA [22]. In the CTAB-SA complex LB film spectrum [figure 3] this carbonyl stretching band is totally absent. The absence of 1710 cm$^{-1}$ band in the spectrum of CTAB-SA complex LB film confirms the attachment of heavy chromophore to the OH group of SA, resulting in the formation of CTAB-SA complex. V. E. Berkheiser et. al. [22] showed that the carbonyl band (1710 cm$^{-1}$) of pure SA disappear when SA interacted with chrysolite [22], indicating the formation of SA-chrysolite complex. The reaction of SA and CTAB is shown below.

$$C_{17}H_{35}COOH + C_{19}H_{42}NBr \rightarrow C_{17}H_{35}COON\ H_{42}\ C_{19} + HBr$$

This is a clear indication that, when the platue region is achieved then interaction between SA and CTAB molecules becomes completed and no SA molecule is left free for further interaction with CTAB molecule.

**UV-Vis absorption and Steady State Fluorescence Spectroscopic Studies of the Complex LB films:**

Rosebengal is an anionic dye whereas CTAB is a cationic surfactant. When RB is mixed with CTAB, the anionic part of RB interacts with the cationic part of the CTAB molecule and thus forms a complex. The interaction scheme is already shown before. This is evidenced from the absorption and fluorescence spectra of pure RB and CTAB-RB mixed solution (figure 4a and 4b). Absorption spectra of pure RB in aqueous solution gives intense absorption peak at 550 nm along with a weak hump at 512 nm, whereas in the RB-CTAB mixed aqueous solution, the high energy band at 512 nm is totally absent and a longer wavelength broad and weak hump in the 570-600 nm region is observed along with the 550 nm intense band. Absence of high energy band at 512 nm and appearance of weak hump in longer wavelength region originate due to the formation of RB-CTAB complex. In figure 4a, absorption spectra of RB microcrystal and RB-CTAB complex microcrystal, are also shown. In RB microcrystal the absorption spectrum has prominent peaks at 526 nm and 562 nm which is red shifted in comparison to the solution absorption spectrum whereas in the RB-CTAB complex microcrystal absorption spectrum, an intense peak is observed at 572 nm alongwith a peak at 528 nm. The shifting of the longer wavelength band at 572 nm confirms the formation of RB-CTAB complex. The absorption spectra of the LB films of RB-CTAB-SA complex are also shown in figure 4a. The films are lifted at various molar ratios of RB and CTAB while the amount of SA is kept fixed. All the films are 10 bilayered.

The most interesting thing observed in the absorption spectra of the complex LB films is that, when the mixture concentration of CTAB:RB is 90:10 and 80:20, they give almost exact replication of the absorption spectra of RB-CTAB complex microcrystal. However, When the RB molar ratio is greater, namely, the volume ratios of CTAB:RB is of the order of 70:30, 60:40, 50:50, the absorption spectra show a totally different pattern. An intense longer wavelength band at 618 nm is observed alongwith a weak band at 530 nm.

This difference in absorption spectra of lower and higher volume ratios of RB may be explained if we look at the absorption spectrum of the CTAB-RB mixed aqueous solution, where a broad weak hump in the 570-600 nm regions is observed. It may be easily concluded that this weak hump is actually consist of two overlapping band system. In the complex LB films, depending on the amount of RB molecules present, two different band systems are developed at different RB concentrations.

The different nature of absorption spectra of complex LB films at higher molar ratios of RB may be due to some molecular level interactions occurring between the CTAB and RB molecules in the LB films. The intimate contact between the molecules due to the organized molecular architecture in the LB films may favour such molecular level interactions. This types of molecular level interactions for ruthenium complexes in the mixed LB films have been suggested by K. Wohnrath et.al. [23] and several others [24, 25]. However for the higher CTAB concentration the required intimate contact for such molecular level interaction is hampered and microcrystalline aggregates are formed in the complex LB films.

It is interesting to mention in this context that, due to change in the packing pattern and orientation of the molecules at the air-water interface as well as also in the films/solid state, the electronic energy level changes and as a result, different band systems may originate.

In figure 4b the fluorescence spectra of the aqueous solution of pure RB, mixed aqueous solution of RB-CTAB complex, RB microcrystal, RB-CTAB complex microcrystal are shown along with the fluorescence spectra of complex LB films at different molar ratios of CTAB and RB. In all the cases of the formation of complex in the Langmuir monolayer, amount of SA is kept fixed.

From the figure it is observed that RB aqueous solution fluorescence spectrum has intense peak at 578 nm and RB-CTAB complex aqueous solution fluorescence spectrum has intense peak at 594 nm. This red shifting of the fluorescence band surely originates due to the formation of RB-CTAB complex. The RB-CTAB microcrystal has intense peak at 604 nm, a shift of about 10 nm in comparison to the RB-CTAB complex aqueous solution. This shifting of fluorescence band may be due to the deformation produced in the electronic energy level of the complex while going from solution to solid state. An intense broad excimer like emission band with peak at around 643 nm originates in the pure RB microcrystal spectrum along with a weak hump at 594 nm. The drastic change between the fluorescence spectra of RB microcrystal and RB-CTAB complex microcrystal may be due to the formation of complex of CTAB and RB molecules which reduces the tendency to form excimer of RB molecules in the solid state.

The most interesting thing is that, in the complex LB films of RB-CTAB-SA at various molar ratios, the fluorescence spectra of RB-CTAB-SA complex LB films show quite interesting features. They possess two band systems with peaks at around 625-675 nm and 560-600 nm regions. The longer wavelength broad band at 625-675 nm region is almost identical to the pure RB microcrystal spectrum indicating the presence of few isolated RB molecules in the complex LB films. Where as the band at 560-600 nm region replicates the CTAB-RB mixed microcrystal spectrum. This certainly concludes that in the complex LB films two types of species exists. In one type CTAB-RB complex is formed and stacked and in another types RB molecules are detached from the CTAB molecules, while CTAB molecules form complex with SA molecules and within the organization of CTAB-SA complex molecules RB molecules get stacked and forms microcrystalline aggregates.

A schematic representation of the LB films structure is shown in figure 5. This is the final modified figure in the light of the discussion of fluorescence spectra.

**Conclusion:**

In conclusion our result shows that a complex Langmuir monolayer is formed at the air-water interface by charge transfer types of interaction between RB doped water soluble CTAB molecules with the SA molecules of a preformed SA Langmuir monolayer. The reaction kinetics and adsorption of RB doped CTAB were observed by monitoring the increase in surface pressure with time while the barrier was kept fixed. FTIR spectra confirms that, when the platue region in the surface pressure versus time graph is achieved then all SA molecules interact with CTAB to form complex. Multilayered LB films of this complex was successfully formed onto quartz substrate. The absorption spectra of complex LB films confirm that due to the organized molecular architecture in the LB films there are some molecular level interactions between CTAB and RB molecules occurred at higher molar ratios of RB. However for the higher CTAB concentration microcrystalline aggregates are formed in the complex LB films. Fluorescence spectroscopic studies reveal that CTAB-RB complex was formed and stacked in the complex LB films. Again few RB molecules were also detached from the CTAB molecules, while CTAB molecules form complex with SA molecules. RB molecules got stacked within the organization of CTAB-SA complex molecules and form microcrystalline aggregates.

**Acknowledgement:**

The authors are grateful to DST and CSIR, Govt. of India for providing financial assistance through FIST-DST Project No. SR/FST/PSI-038/2002 and CSIR project Ref. No. 03(1080)/06/EMR-II

**References:**
1. S. Deb, S. Biswas, S.A. Hussain, D. Bhattacharjee, Chemical Physics Letters 405 (2005) 323
2. S. A. Hussain, S. Deb, S. Biswas and D. Bhattacharjee, Spectrochim. Acta A 61 (2005) 2448
3. S.A. Hussain, S. Deb, D. Bhattacharjee, J. Lumin. 114 (2005) 197
4. S. Biswas, S. A. Hussain, S. Deb, R. K. Nath, D. Bhattacharjee, Spectrochim. Acta A 65 (2006) 628
5. M. Kawaguchi, M. Yamamoto, T. Kato, Langmuir 14 (1998) 2582
6. J. Engelking, H. Menzel, Eur. Phys. J. E 5 (2001) 87
7. J. Engelking, D. Ulbrich, H. Menzel, Macromolecules 33 (2000) 9026
8. D. Vollhardt, V. Melzer, J. Phys. Chem. B 101 (1997) 3370
9. D. Vollhardt, V. Melzer, V. Fainerman, Thin Solid Films 327-329 (1998) 842
10. D. Vollhardt, Advances in Colloid Surface Science 79 (1999) 19
11. Z. Kozarac, A. Dhathathreyan, D. Mobius, Coll. Surf. 33 (1988) 11
12. H. Dautzenberg, W. Jaeger, J. Kotz, B. Phillip, C. Seidel, D. Stscherbina, Polyelectrolytes-Formation, Characterization, Application, carl hanser Verlag, Munich, 1994
13. H. Waltve, C. Harrats, P. Muller Buchmann, R. Jerome, M. Stamm, Langmuir 15 (1999) 1260
14. N. A. Kotov, I. Dekany, J. H. Fendler, Adv. Matter 8 (1996) 637
15. K. Miyano, K. Asano, M. Shimomura, Langmuir 7 (1991) 444
16. K. Asano, K. Miyano, H. Uki, M. Shimomura, Y. Ohta, Langmuir 9 (1993) 3587

17. R. Volinsky, S. Kolusheva, A. Berman, R. Jelinek, Biochim. Biophys. Acta, Article in press
18. P. Fromherz, Biochim Biophys. Acta 225 (19971) 382
19. F. Ebler, Dissertation Universitat Mainz, 1998
20. S. Sundaram, J. K. Ferri, D. Vollhardt, K. J. Stebe, Langmuir 14 (1998) 1208
21. G. Xiong, Z. Zhi, X. Yang, L. Lu, X. Wang, J. Material Science Letters 16 (1997) 1064
22. V. E. Berkheiser Cáliz and Cáliz Minerals 30 (1982) 91
23. 22. K. Wohnrath, L. R. Dinelli, S. V. Mello, C. J. L. Constantino, R. M. Leblanc, A. A. Batista, O. N. Oliveira, Jr. J Nanoscience Nanotechnology X (2005) 1
24. K. Wohnrath, C. J. L. Constantino, P. A. Antunes, P. M. dos Santos, A. A. Batista, R. F. Aroca, O. N. Oliveira, Jr. J. Phys. Chem. B 109 (2005) 4959
25. M. Ferreira, K. Wohnrath, A. Riul Jr., J. A. Giacometti, O. N. Oliveira Jr. J. Phys. Chem. B 106 (2002) 7272

**Figure captions:**

Figure-1: Schematic representation of the interaction scheme of RB-CTAB complex with the SA molecules of the preformed SA monolayer at the air-water interface.

Figure 2a: Surface pressure versus time graph of the RB-CTAB-SA complex monolayer with various amount of RB-CTAB mixture spread on the water for initial starting surface pressure at 0.5 mN/m. Numbers denote the corresponding volume ratios of CTAB and RB mixture.

Figure 2b: Surface pressure versus time graph of the RB-CTAB-SA complex monolayer with various amount of RB-CTAB mixture spread on the water for initial starting surface pressure at 15 mN/m. Numbers denote the corresponding volume ratios of CTAB and RB mixture.

Figure 3: FTIR spectra of pure SA and CTAB-SA complex LB films on silicon substrate.

Figure 4a: UV-Vis absorption spectra of RB-CTAB-SA complex LB films with various amount of RB-CTAB mixture along with the RB, RB-CTAB solution, RB and RB-CTAB microcrystal spectra. Numbers denote the corresponding volume ratios of CTAB and RB mixture.

Figure 4b: Fluorescence spectra of RB-CTAB-SA complex LB films with various amount of RB-CTAB mixture along with the RB, RB-CTAB solution, RB and RB-CTAB microcrystal spectra. Numbers denote the corresponding volume ratios of CTAB and RB mixture.

Figure 5: Schematic representation of RB-CTAB-SA complex Langmuir monolayer at the air-water interface.

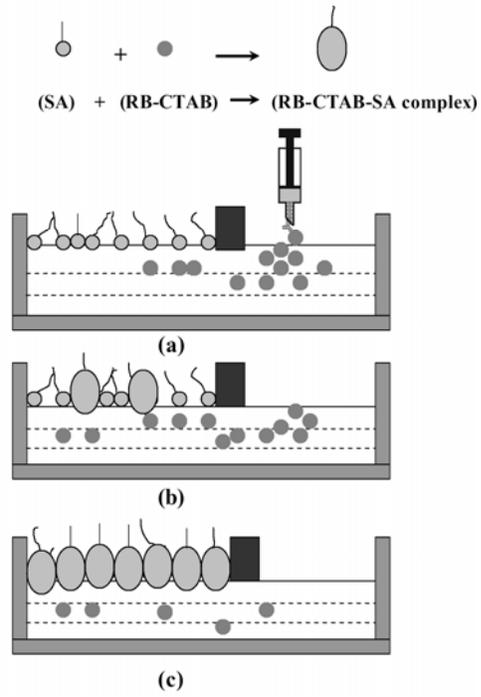

Figure 1: S. Biswas et. al.

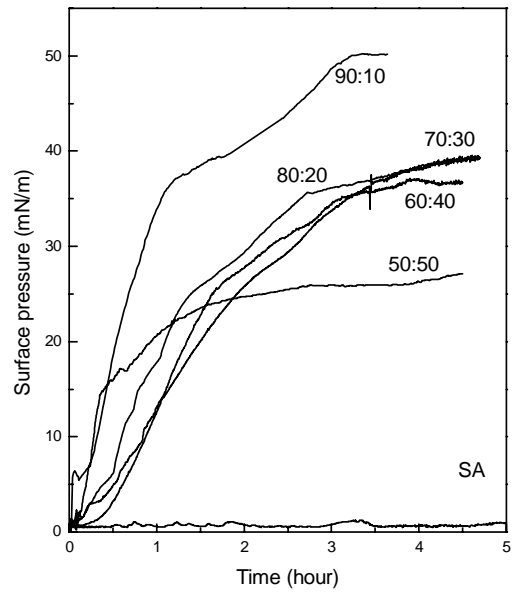

Figure 2a: S. Biswas et. al.

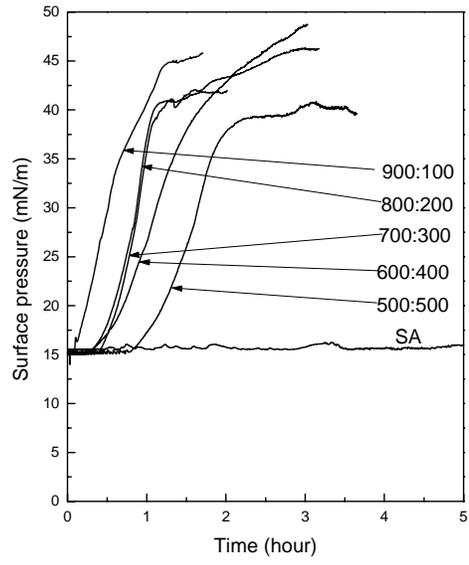

Figure 2b: S. Biswas et. al.

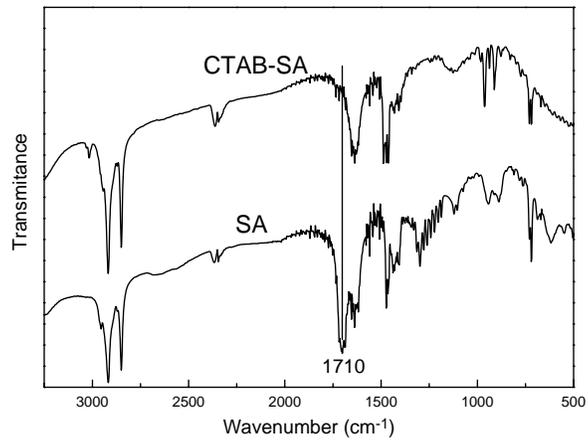

Figure 3 S. Biswas et. al.

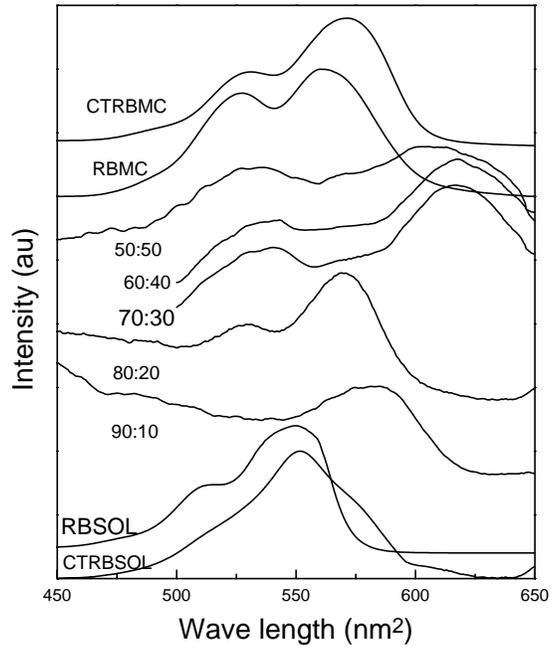

Figure 4a: S. Biswas et. al.

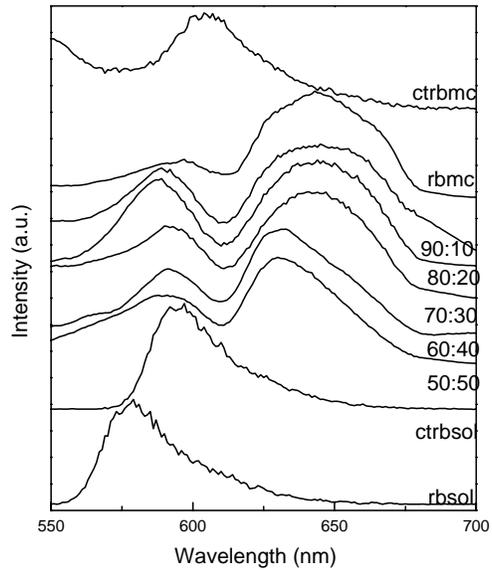

Figure 4b: S. Biswas et. al.

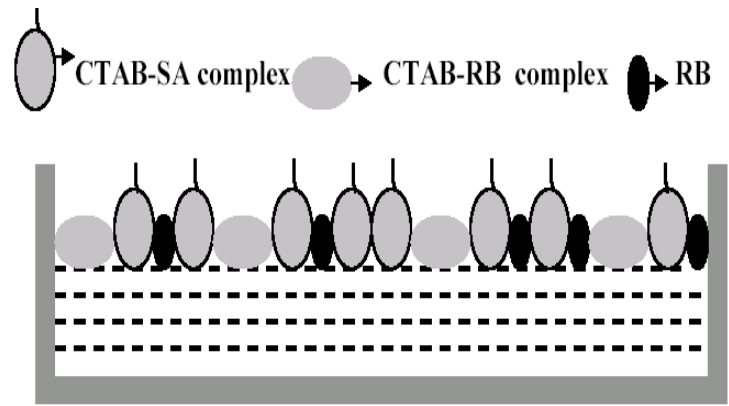

Figure 5: S. Biswas et. al.